\documentclass{ws-procs975x65}

\def\beq{\begin{equation}}
\def\eeq{\end{equation}}

\begin{document}

\title{Ultraluminous X-ray sources as super-Eddington accretion disks}
\author{Sergei Fabrika$^{1,2*}$, Alexander Vinokurov$^1$, Kirill Atapin$^1$}

\address{$^1$Special Astrophysical Observatory \\
Nizhnij Arkhyz, 369167, Russia \\
$^2$Kazan Federal University, 420008 Kazan, Russia\\
$^*$E-mail: fabrika@sao.ru\\
}

\begin{abstract}
The origin of Ultraluminous X-ray sources (ULXs) in external galaxies whose
X-ray luminosities exceed those of the brightest black holes in our Galaxy
by hundreds and thousands of times is mysterious. The most popular models
for the ULXs involve either intermediate mass black holes (IMBHs) or 
stellar-mass black holes 
accreting at super-Eddington rates. Here we review the ULX properties, their
X-ray spectra indicate a presence of hot winds in their accretion disks
supposing the supercritical accretion. However, the strongest evidences come
from optical spectroscopy. The spectra of the ULX counterparts are very
similar to that of SS\,433, the only known supercritical accretor in our
Galaxy.
\end{abstract}


\bodymatter
\vspace*{0.5cm}

Ultraluminous X-ray sources (ULXs) are X-ray sources with luminosities
exceeding the Eddington limit for a typical stellar-mass black hole 
$\sim 2 \times 10^{39}$ erg s$^{-1}$. Despite their importance 
in understanding the origin of supermassive black holes that reside 
in most of present galaxies, the basic nature of ULXs remains unsolved.\cite{feng11} The most popular
models for the ULXs involve either intermediate mass black holes (IMBH,
$10^3$--$10^4$\,M$_{\odot}$)\cite{madau01} with standard accretion disks or 
stellar-mass black holes ($\sim 10$\,M$_{\odot}$) accreting at super-Eddington 
rates. The last idea has been suggested\cite{fabr01} because of an analogy 
with SS\,433, the only known super-accretor in the Galaxy,\cite{fabr04} 
and developed in (Ref. 13, 18). It was proposed that SS\,433 supercritical 
disk's funnel being observed nearly face-on will appear as extremely bright 
X-ray source. Both scenarios, however, require a massive donor in a close binary. 

Most of the ULXs are associated with the star-forming regions and surrounded 
with nebulae of a complex shape, indicating a dynamical influence of the 
black hole.\cite{abol07} They are not distributed throughout galaxies as it
would be expected for IMBHs originating from low-metallicity Population
III stars. The IMBHs may be produced in a runaway merging in a core of
young clusters. In this case, they should stay within the
clusters. It has been found\cite{pout13} that all brightest X-ray sources in
Antennae galaxies are located nearby to very young stellar
clusters. It was concluded that the sources were ejected in the
process of formation of stellar clusters in the dynamical few-body
encounters and that the majority of ULXs are massive X-ray binaries with
the progenitor masses larger than 50\,M$_{\odot}$.

The X-ray spectra of the ULXs often show a high-energy curvature with 
a downturn between $\sim 4$ and $\sim 7$\,keV. It was called ``ultraliminous 
state''.\cite{gladst09} The curvature hints that the ULX accretion disks 
are not standard. Inner parts of the disks may be covered with hot outflow 
or optically thick corona, which Comptonizes the inner disk photons.

\begin{figure}[ht]
\includegraphics[width=5.5in]{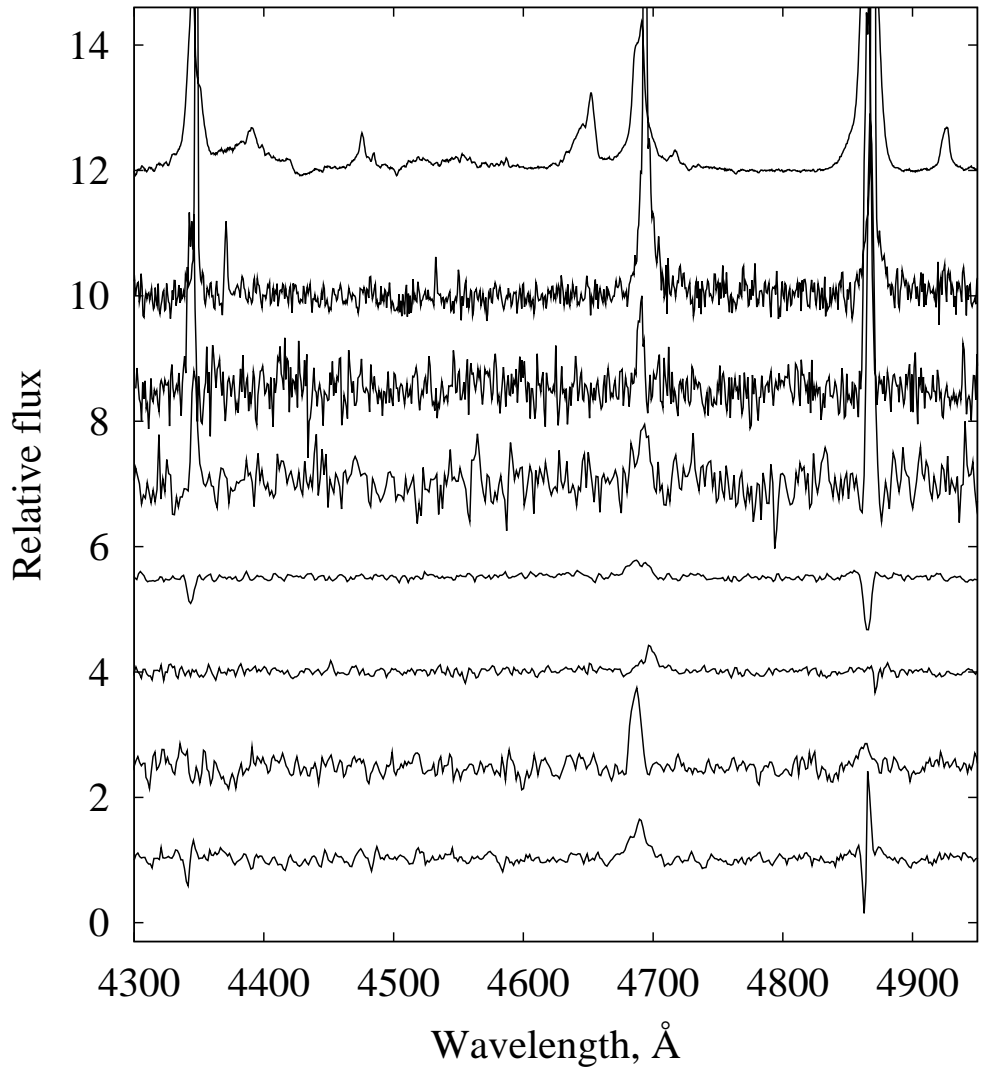}
\caption{Normalized optical spectra of ULX counterparts. From top to bottom: SS\,433 (1),
NGC\,5408 X-1 (2), NGC\,4395 X-1 (3), NGC\,1313 X-2 (2), NGC\,5204 X-1, NGC\,4559 X-7, 
Holmberg\,IX X-1 and Holmberg\,II X-1 (1). The numbers in brackets mean optical telescopes
1 -- Subaru telescope, 2 -- VLT (ESO), 3 -- Russian BTA telescope. The spectra are very 
similar one another, they may represent rare type of massive stars WNL\citep{crowther11} 
or LBV stars in their hot states.\citep{shol11,shol15} All the spectra are also similar 
to SS\,433.\citep{kubota10} This means that the spectra of the ULX counterparts are formed 
in hot winds.}
\label{aba:fig1}
\end{figure}

Spectra of almost all optical counterparts of studied ULXs (with SS\,433 included) 
are shown in Fig.\,1. 
Main features in all the spectra are the bright He\,II\,$\lambda 4686$, hydrogen 
H$\beta$, and H$\gamma$ emission lines. The lines are obviously broad; the widths
range from 500 to 1500\,km s$^{-1}$. Some hydrogen lines are contaminated by nebulae
(over-subsraction of narrow absorptions) associated with the ULXs. However the broad emission components 
in H$\beta$ are clearly seen. In the SS\,433 spectrum, besides moving lines which 
form in the relativistic jets, 'stationary' emission lines, the Bowen blend 
CIII/NIII $\lambda \lambda 4634 - 4651$ and three He\,I lines are observed. 
Some of these lines are seen in the ULX spectra.\citep{fabr15}

All the spectra of the ULXs 
(the spectra were reduced by us) are surprisingly similar to one another. 
The optical spectra are also similar to that of SS 433, although the ULX spectra 
indicate a higher wind temperature. It was  suggested in (Ref. 6) that 
the ULXs must constitute a homogeneous class of objects, which most likely 
have supercritical accretion disks. 

\begin{figure}[ht]
\begin{center}
\includegraphics[width=3.3in]{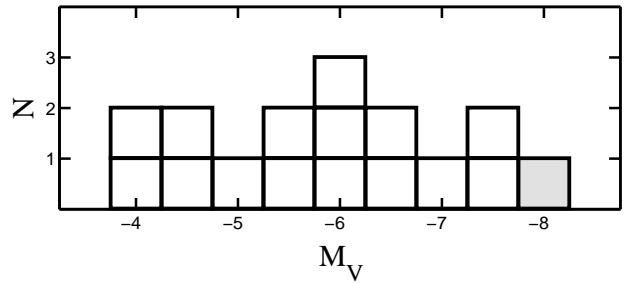}
\end{center}
\caption{
Absolute magnitudes of all well-studied ULXs and SS\,433 (shadowed). The data are 
from (Ref. 6) with some updates from (Ref. 9, 18).
}
\label{aba:fig2}
\end{figure}

\begin{figure}[ht]%
\parbox{2.6in}{\includegraphics[width=2.7in]{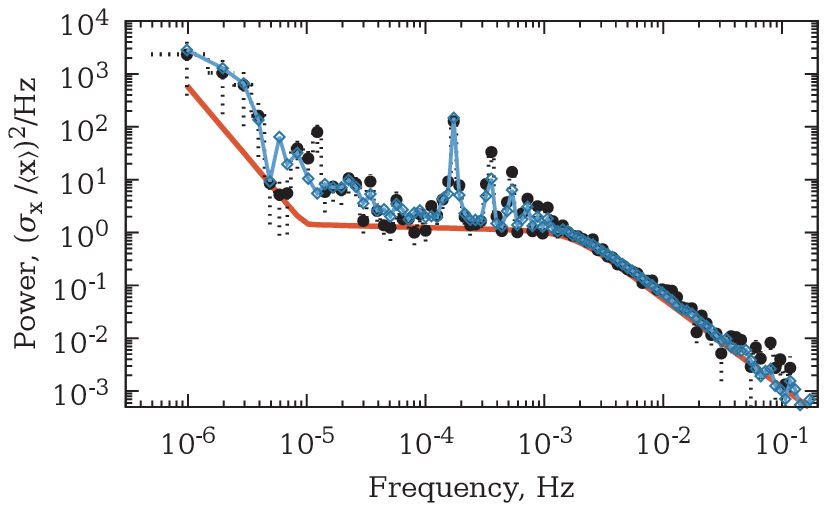}}
\parbox{2.1in}{\includegraphics[width=2.7in]{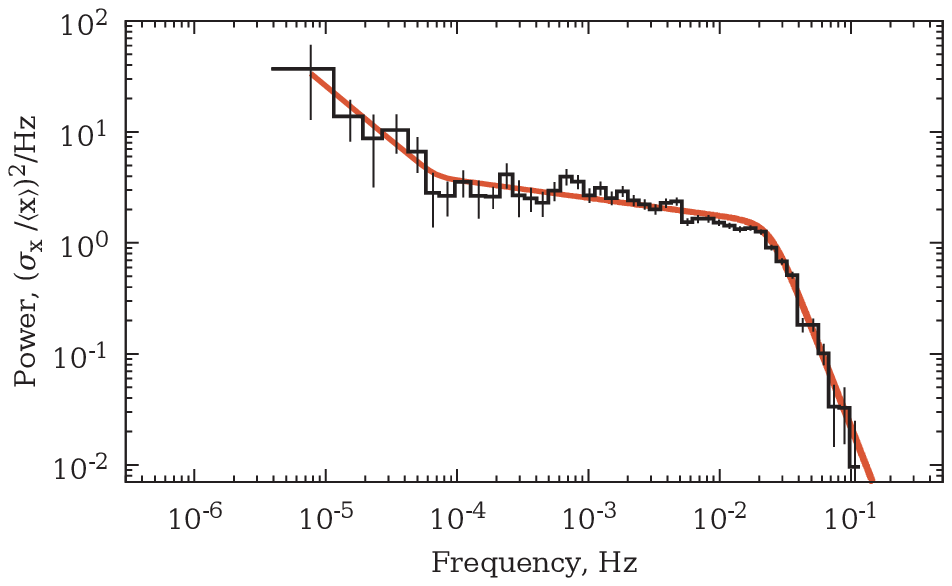}}
\caption{Power density spectra of SS\,433 (left) and the ULX NGC\,5548 X-1 
(right). In SS\,433 we observe a flat portion\citep{atap15} which may be
considered as alpha-viscosity fluctuations appearing at the spherization radius
as it was originally proposed\citep{lyubar97} for standard disks.\citep{SS73}
The power spectrum of SS\,433 has been obtained from a single ten-days ASCA
observation. The circles and dotted line are the observed power sectrum, the
red (dark grey) solid line is initial model of the accretion disk intrinsic
variability. The blue (light grey) line and diamonds are Monte Carlo model,
which takes into account gaps in the observation and extra variability added
by eclipse occurred during this observations. The power spectrum of SS\,433 has 
a flat part stretching from $\sim 10^{-5}$ to $\sim 10^{-3}$~Hz.
The power spectrum of NGC\,5408 X-1 has been obtained by averaging of six most long
observations from XMM-Newton. A model with two breaks fitting the spectrum
is shown by solid line. This object as well has a flat part in the power spectrum.
}
\label{fig3}
\end{figure}

The total luminosity of a supercritical disk is proportional to the 
Eddington luminosity with an additional logarithmical factor depending on
the original mass accretion rate,\cite{SS73,pout07} because 
the excess gas is expelled 
as a disk wind and the accreted gas is advected with the photon 
trapping, contributing little to the photon luminosity. 
However, the UV and optical luminosity in such disks may strongly 
depend on the original mass accretion rate, because these budgets are mainly 
produced by the reprocess of the strong irradiation from the  
wind (the excess gas). Optical spectra of SS\,433 and the ULX
counterpart are nearly the same, but in X-rays they are drastically 
different because we cannot observe the funnel in SS\,433 directly.
It was found\cite{fabr15} that the mass accretion rates in the ULXs 
may be by a factor of 1.5--6 smaller and their wind temperatures are 
by 1.4--4 times higher than those in SS\,433. 
In Fig.\,2 we show absolute magnitude of all well 
studied ULXs together with SS\,433. Thus, one may interpret 
that SS\,433 is intrinsically the same as ULXs but an extreme case 
with a particularly high mass accretion rate, which could explain the 
presence of its persistent jets.\cite{fabr04}

In Fig.\,3 we present one more evidence of the super--Eddington accretion 
in ULXs. The power density spectrum of SS\,433 exhibit a flat part in 
the $10^{-5} - 2\times 10^{-3}$\,Hz frequency range.\cite{atap15} The 
presence of such a part is related to the abrupt change in the disk 
structure and the viscous time at the spherization radius. In this 
place the accretion disk becomes thick, which reduces drastically 
the time of passage of matter through the disk.\cite{SS73,lyubar97}
The same picture is observed in the well-studied ULX NGC\,5408 X-1. 
We need longer observations of the ULXs to study their power density 
spectra in more details. 

\section*{Acknowledgments}

The research was supported by the Russian Scientific Foundation grant (N\,14-50-00043)
and RFBR grants (N\,15-42-02573, 16-02-00567, 16-32-00210). SF acknowledges support of 
the Russian Government Program of Competitive Growth of Kazan Federal University.

\end{document}